\documentclass[12pt,preprint]{aastex}
\usepackage{apjfonts,emulateapj5}

\slugcomment{}
\shorttitle{Speedy Clouds in the NLR of NGC 1068}
\shortauthors{Cecil et al.}

\begin{document}

\title{Spatial Resolution of High-Velocity Filaments in the Narrow-Line
Region of \\
NGC 1068: Associated Absorbers Caught in Emission?}

\author{Gerald Cecil}

\affil{Dept. of Physics \& Astronomy, U. of N. Carolina, Chapel Hill, NC
27599-3255}

\email{gerald@thececils.org}

\author{Michael A. Dopita, Brent Groves}

\affil{Institute for Advanced Studies, Australian National University, Canberra
ACT, Australia}

\author{Andrew S. Wilson}

\affil{Dept. of Astronomy, U. of Maryland, College Park, MD 20742}

\author{Pierre Ferruit, Emmanuel P\'econtal}

\affil{Observatoire de Lyon, France}

\and{}

\author{Luc Binette}

\affil{UNAM, Mexico}

\begin{abstract}
Using the STIS spectrograph on \emph{HST} we have obtained a grid of 
[\ion{O}{3}]$\lambda\lambda$4959,5007 and H$\beta$ emission-line
spectra at $0\farcs05\times0\farcs19$\arcsec\ and 60 km~s$^{-1}$ (FWHM) 
resolution that covers much of the NLR of NGC 1068.
We find emitting knots that have blueshifted radial velocities up to
3200 km s\( ^{-1} \) relative to galaxy systemic, are
\( 70-150 \) pc NE of the nucleus and up to 40
pc from the radio jet, emit several percent of the NLR line flux but
no significant continuum,
span a small fraction of the sky as seen from the nucleus, coincide
with a region of enhanced IR coronal-line emission,
show gradients in radial velocities of up to 2000
km~s\( ^{-1} \) in 7 pc, span velocity extents averaged over \( 0\farcs 1\times 0\farcs 2 \)
regions of up to \( 1250 \) km~s\( ^{-1} \), have ionization parameter
\( \mathcal{U} \ga 0.1 \), and ionized masses \( \sim 200 \) M\( _{\odot }/n_{e,4} \)
(\( n_{e,4}=10^{4} \) cm\( ^{-3} \)). 
The brightest parts of the blueshifted knots are often
kinematically contiguous with more massive
clouds nearer the jet that are moving with velocities
of \( \leq 1300 \) km s\( ^{-1} \) relative to galaxy 
systemic. However, some knots at \( 1\farcs 5-2\farcs 5 \) radii appear
as bright points in a broken shell of radius \( \sim 0\farcs 55 \)
(40 pc) that is expanding at up to 1500 km~s\( ^{-1} \), implying
a dynamical age of \( \sim 1.3\times 10^{4} \) yrs.
Between 2\farcs5--4\farcs5 from the nucleus, emission is
redshifted relative to systemic, a pattern that we interpret as
gas in the galaxy disk being pushed away from us by the NE radio lobe.
We argue that the blueshifted knots are ablata from disintegrating
molecular clouds that are being photoionized by the AGN, and are being
accelerated radiatively by the AGN or mechanically by the radio jet.
In their kinematic properties, the knots
resemble the associated absorbers seen projected on the UV continua
of some AGN.
\end{abstract}


\section{Introduction}

Emission-line profiles that extend over several thousand km~s\( ^{-1} \)
are hallmarks of activity in galaxy nuclei of all luminosities. Some
AGN also show absorption lines at UV rest wavelengths, originating from warm
gas located outside the region that generates the nonstellar continuum.
This gas
is invariably blueshifted over a range of several thousand km~s\( ^{-1} \)
from the galaxy systemic velocity, yet often shows discrete components
that each span several hundred km~s\( ^{-1} \) \citep{ha97}. 
What process accelerates the clouds?

Spatially complete spectral maps of nearby active galaxies
can separate clouds that are agitated mechanically by the AGN from
those that are simply illuminated by AGN photons (e.g.\ undisturbed gas in an
{}``ionization cone'').
In this paper we report on observations made with the Space Telescope
Imaging Spectrograph (STIS) and its medium-resolution (M)
gratings to map the ``Narrow-Line Region" (NLR) of the Seyfert galaxy
NGC 1068.
In this NLR, \cite{al83} sampled a cloud complex that spans \( \sim 200 \)
pc, is centered \( \sim 210 \) km~s\( ^{-1} \) blueward of galaxy
systemic velocity, and has \( \sim 1670 \) km~s\( ^{-1} \) FWHM
(\( \sim 2200 \) km~s\( ^{-1} \) according to the \emph{HST} FOS
spectra of \cite{ca91}) velocity extent. The complex of clouds discovered
by \cite{wa68}, and mapped in detail by 
\citet[CBT hereafter]{ce90} and \cite{ar96} (see also \citet{ba87}), 
was shown from FOS spectra \citep{kr98} to be mostly
photoionized by nuclear radiation, and to
have bulk motion consistent with acceleration that is independent of radius
\citep[CKb hereafter]{CK00}. A recent STIS spectrum 
\citep[KC1 hereafter]{kr00}
shows that emission-line centroids near the {}``continuum
hotspot'' (a feature with substantial scattered nuclear light located
\( \sim 0\farcs 17 \) {[}12 pc{]} N of the optically obscured nucleus)
exhibit a trend towards greater blueshifted velocities as the ionization 
potential
of the line species increases, a result that extends the
earlier finding by \cite{mo96} who found that coronal, near-IR lines
are extended across a \( \leq 4\arcsec  \) region. 

Section 2 describes the acquisition and reduction of the STIS spectra.
We show in \S3 that the fastest knots have been accelerated to radial velocities
blueshifted from galaxy systemic by $>$3200 km~s\( ^{-1} \), a pattern
not evident in the single STIS spectrum discussed by KC1 and CKb. We consider
various mechanisms capable of accelerating clouds and knots in \S4, outline
why the knots may be related to quasar ``associated absorbers" in \S5, 
briefly consider
future observations to test this hypothesis in \S6, and summarize
our conclusions in \S7.
In this paper, we assume a distance to NGC 1068 
of 14.4 Mpc, so 1\arcsec = 70 pc.

\section{STIS Long-Slit Spectra}

\subsection{Data Acquisition}

Deep spectra at high spatial resolution are necessary to map
and characterize the various dynamical subsystems in this NLR. We
have therefore used the STIS, splitting 14 \emph{HST}
orbits evenly between M-grating
(R $\sim$ 5,000) maps of the {[}\ion{O}{3}{]} and H\( \beta  \) emission-line
profiles, and less extensive L-grating (R $\sim$ 1,000) maps of far- and
near-UV line fluxes. We used the \( 0\farcs 2 \)-wide slit
(\( 0\farcs 19 \) on sky) as a reasonable compromise between resolution
and mapping efficiency. To maximize sensitivity, 
we binned pairs of pixels along the axis of wavelength dispersion 
during CCD readout of the M-grating spectra, yielding
60 km~s\( ^{-1} \) FWHM resolution. 

We planned to obtain spectra both along and perpendicular to the radio
jet. However, available guidestars restricted us to P.A.\ 38\arcdeg,
close to the axis of the large-scale radio jet. After a standard STIS
peakup on the nucleus, we performed a small-angle offset perpendicular
to the slit for each map position. We obtained five long-slit spectra
before a safe event delayed the next spacecraft visit sufficiently
that the large changes in the orbit of \emph{HST} prevented us from reusing
target offsets from guidestars. This required us to spend an unanticipated
half-orbit a year later to reacquire guidestars. The added overhead,
together with a slight positional shift between the two pointings,
meant that we obtained a total of six spectra with parallel slits that cover
$\sim$2/3 of the area of the NLR, with spatial resolution 0\farcs05 
along the slit (and jet axis) and 0\farcs19 in the perpendicular direction,
see Fig.\ \ref{fig:fig1}.
We supplemented these data with several G430L and
G430M spectra from the \emph{HST} archive (P.I.\ R.\ Antonucci) that were unbinned
in wavelength and used the 0\farcs1-wide slit; the 
three archival M-grating exposures yielded a single
spectrum along P.A. 10\arcdeg\ that intersects the continuum hotspot,
close to the P.A. 22\arcdeg\ orientation used by a published STIS spectrum
(KC1).
Standard STIS processing with on-the-fly recalibration using optimal
files produced wavelength and flux calibrated spectra whose parameters
are listed in Table 1.

\placetable{Table 1}

\subsection{Data Reduction}

We used the IRAF `cosmic' task to remove charge spots and hot pixels.
To derive line fluxes,
we used the average reddening derived by KC1 for the blue-wing of
the line profiles, \( E_{\textrm{B--V}}=0.35 \) as an initial guess
to be refined at different points along the slits during modeling
of our more dust-sensitive UV spectra \citep{gr01}.
This reddening is consistent
with estimates from the {[}\ion{Fe}{2}{]}\( \lambda  \)1.257 \( \mu  \)m
to 1.64 \( \mu  \)m flux ratios across the NLR \citep{mo96}.

The spectra were examined for instrumental scattered light following
a procedure similar to that of \cite{Ne00}; none was found. After
masking emission lines, we used the IRAF `continuum' task with third-order
Chebyshev polynomials to model and subtract continuum light on a row
by row basis.

\begin{figure}
\epsscale{0.6}
\plotone{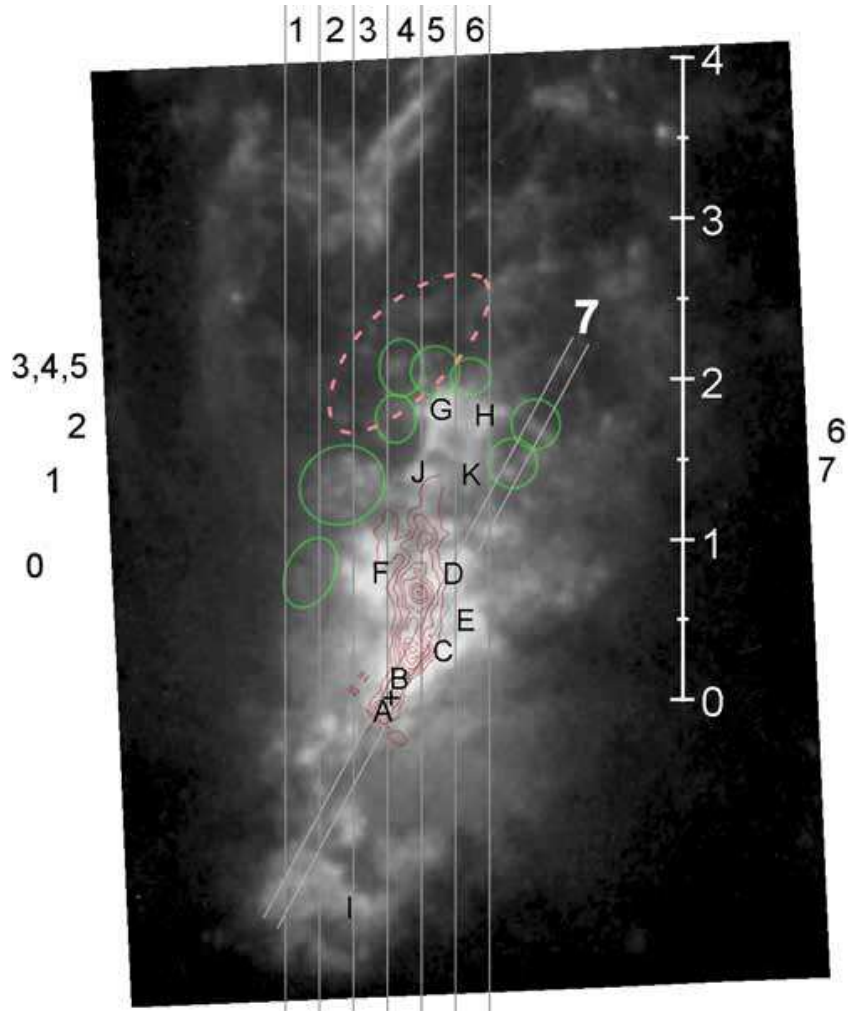}
\caption{\label{fig:fig1}FOC image of [\ion{O}{3}]$\lambda\lambda$4959,5007 
line emission across the circumnuclear region of NGC 1068
\citep{ma94}.  The greyscale is proportional to the log of
the intensity. The vertical direction is in P.A.\ 38\arcdeg, with NE at top.
denote the high-velocity regions (HV0--HV7) described in \S3.1.2; they are also
drawn in Fig.\ \ref{fig:fig2}
to define the relationship between this image and the
locations of the slits. The numbers 0--7 of these clouds are given outside
the figure at left and
right.  The vertical scale (white line and vertical numbers)
originates at the (obscured) nucleus (which coincides with radio
source S1, denoted by +) and is in arcseconds.  The numbers at top and the
white 7 in the figure refer to the slits, the boundaries of which are 
shown as vertical and diagonal
lines. The spectra from these 7 slit positions are shown
in Fig.\ \ref{fig:fig2}.
The location of the 
Doppler ellipsoid is dashed in magenta. The V-shaped feature discussed
in \S3.1.4 and \S\ref{vshape} is at top, between 3 and 4\arcsec\ radius.}
\end{figure}

\placefigure{Fig. \ref{fig:fig1}}

From ground-based spectral maps \citep[for example CBT, and][]{ar96}, 
line emission in this NLR is
known to be structured into roughly a half-dozen major subsystems
with different kinematics, excitation, and geometric distribution.
The gas distribution appears to be even more complex
in the \emph{HST} Faint Object Camera (FOC) image (Fig. \ref{fig:fig1}), where
many knots are unresolved spatially. This implies that there will
be a {}``slit effect'' on derived radial velocities, the magnitude of which
depends on where the knot falls within our \( 0\farcs 19 \)-wide
slit.  This effect is small
compared to the line widths of features in
our data, and amount to \( \pm 62 \) km~s\( ^{-1} \) (\( \pm  \)1
resolution element) per maximum \( \pm 0\farcs 095 \) displacement
(STIS Instrument Handbook, at \url{http://www.stsci.edu/instruments/stis}).

The lines of the {[}\ion{O}{3}{]}$\lambda\lambda$4959,5007
doublet are often blended because
of the large range of gas velocities. To clean line profiles
for analysis, we therefore used two complementary approaches: 1) parameterize
each line with up to 25 Gaussians and constrain the two lines to
have the correct theoretical ratio where they overlap, 2) multiply the profile
of the
red wing of {[}\ion{O}{3}{]}\( \lambda 5007 \) by the theoretical 
$\lambda$4959/5007 flux ratio and then subtract the result from the
observed red wing of the \( \lambda 4959 \) line. We synthesized
the {[}\ion{O}{3}{]}-only image from our spectral grid, and then aligned this
image to the FOC image to better than \( 0\farcs 1 \)
accuracy (half of the slitwidth); we thus verified that the FOC F509N filter
passes all velocity components in our spectra. \cite{ca97} have performed
absolute astrometry and aligned the
FOC image with VLA radio images, finding that radio feature S1 coincides with 
the warped maser disk, presumably the AGN, while S2 is a knot in the jet.

\section{Empirical Results}

\subsection{Kinematic Components in the {[}\ion{O}{3}{]} and H\protect\( \beta \protect \)
Emission-Line Profiles}

\subsubsection{Overview}

Our STIS spectra resolve the spatio-kinematic structure of this NLR,
subdividing the major complexes into three major kinematic components that
have different velocity substructure yet similar 
[\ion{O}{3}] and H$\beta$ line profiles:
\begin{enumerate}
\item that associated with the rotating galaxy and its ionization cone 
\cite[the narrow ``spike'' in the NE quadrant, see][]{al83}, 
\item redshifted filaments, mostly associated with the NE radio lobe,
which arise as the lobe pushes down on the disk, as originally suggested by
\citet[PFBW hereafter]{pe97},
\item blueshifted components that, at fainter levels, extend to velocities 
of more than 3000 km~s$^{-1}$ from systemic
(larger than the 2950 km~s$^{-1}$ separation of the [\ion{O}{3}] doublet).
\end{enumerate}

\begin{figure}
\epsscale{0.95}
\plotone{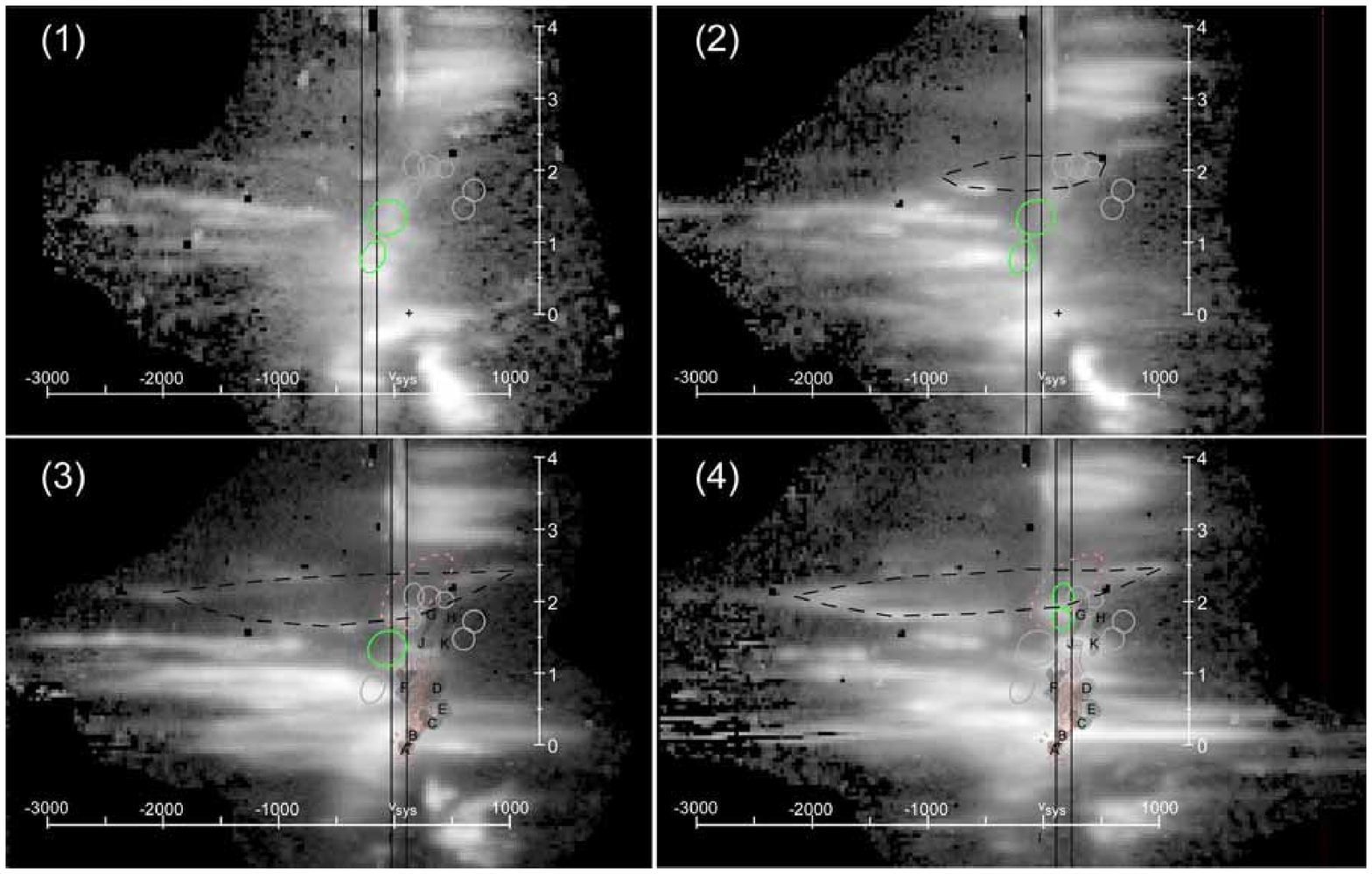}
\plotone{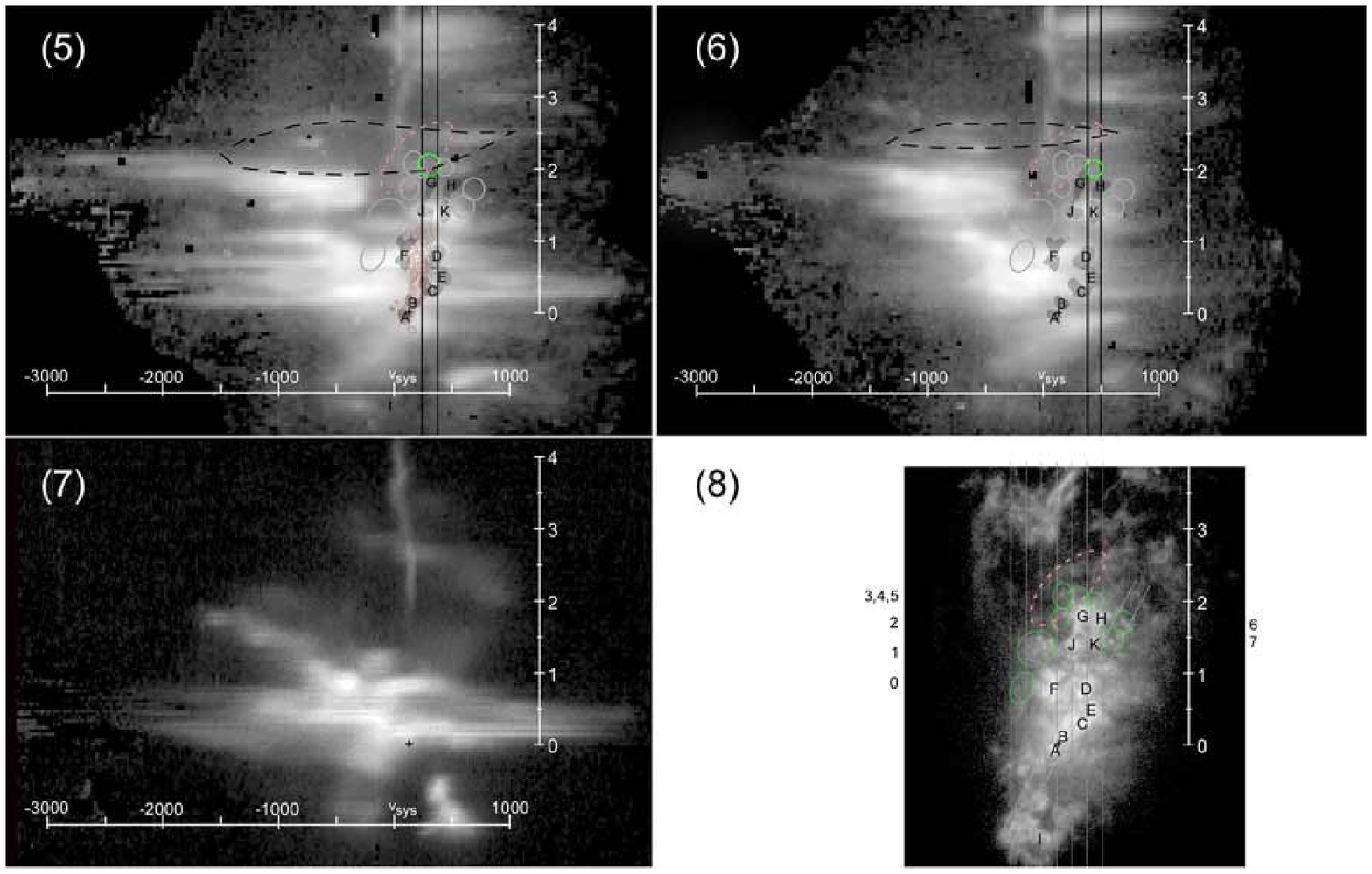}
\caption{\label{fig:fig2}Continuum-subtracted, log-scaled intensity, 
spatial-spectral 
images of {[}\ion{O}{3}{]}$\lambda5007$ emission-line profiles, with 
position along the slit vertical and 
wavelength horizontal. The $\lambda4959$ profile has been subtracted.
Slit P.A. is 38\arcdeg, except for panel (7) where
it is P.A. 10\arcdeg\ (marked by the parallel lines in panel 8, which repeats 
the FOC image shown in Fig.\ \ref{fig:fig1}).
The archival spectrum in panel (7) has been binned spatially to resemble 
spectra in panels (1--6).
The position of the slit w.r.t. the circles and ellipses denoting the
high-velocity (HV) complexes (see Fig.\ \ref{fig:fig1} and Table 2) is shown in panels 
(1--6).
The vertical scale is in arc-seconds from the nucleus (radio source S1)
and the horizontal scale is in km~s$^{-1}$ relative to the galaxy systemic value.
Note the very broad ($>$5,000 km~s$^{-1}$ FWZI) [\ion{O}{3}] profiles at knot 
B in slit 4. The Doppler ellipsoid (\S\ref{sec:shocks}), which is dashed in 
magenta in Fig.\ \ref{fig:fig1},
has its kinematic manifestation marked by a black dashed oval in panels 2--6.
A high-fidelity, color version of this figure is at
http://www.thececils.org/science/n1068/spectra.jpg}
\end{figure}

\begin{figure}
\plotone{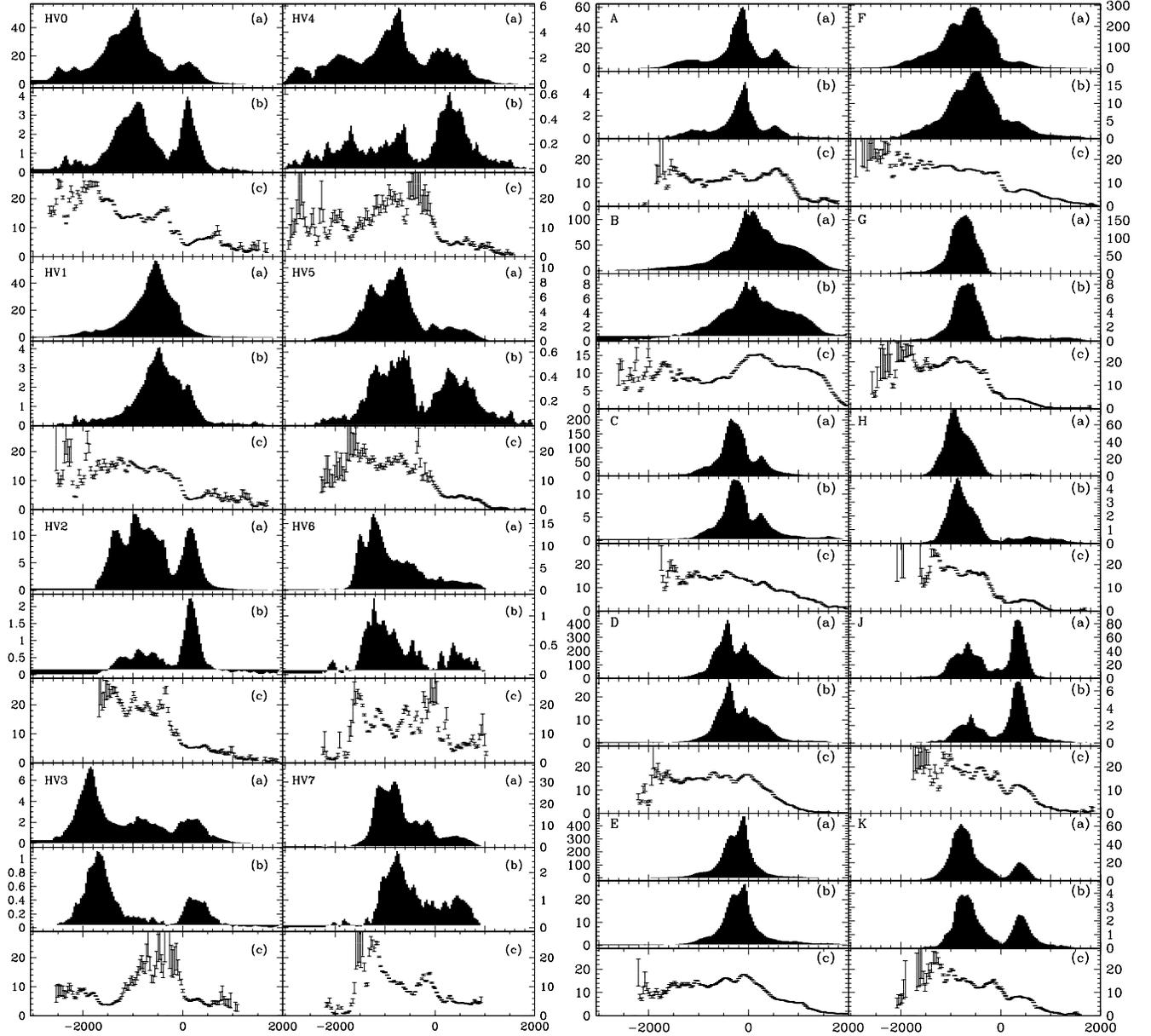}
\caption{\label{fig:f12}Shown are the dereddened line profiles of the 
high-velocity knots (HV0--HV7: left diagram) and more
massive clouds (A--K: right diagram).
In each set of three spectra, the top profile (a) is {[}\ion{O}{3}{]}$\lambda5007$ after contamination by $\lambda$4959 has been removed,
the middle (b) is H\protect\( \beta \protect \), and the bottom (c) shows the
range of the ratio [\ion{O}{3}]$\lambda$5007/H$\beta$ at each velocity within
$\pm1\sigma$ statistical
plus estimated systematic uncertainties from the continuum subtraction.
Velocities are relative to the galaxy systemic value. 
The spectral resolution is 60 km~s$^{-1}$ FWHM for the [\ion{O}{3}] and 
H$\beta$ profiles; the line ratio plots have been smoothed with a boxcar of width 
160 km~s$^{-1}$.
The vertical scale is
in units of \protect\( 10^{-18}\protect \) ergs~s\protect\( ^{-1}\protect \)~
cm\protect\( ^{-2}\protect \) pixel$^{-1}$, where the pixel bin width is 
33 km~s$^{-1}$.}
\end{figure}

We have associated kinematic components with
individual clouds or cloud complexes located in the FOC image
(Fig.~\ref{fig:fig1}).
Slit numbers are indicated in Fig.\ 1 and also
refer to panel numbers in the spectrograms of the seven slit positions 
Fig.~\ref{fig:fig2}.
We focus on the various blueshifted components, summarizing
their line profiles and the 
velocity dependences of their [\ion{O}{3}]/H$\beta$ flux ratios
in Fig.~\ref{fig:f12}. 
All velocities are quoted relative to the systemic value of
the galaxy \cite[1148 km s\( ^{-1} \),][]{br97}.

\placefigure{Fig. \ref{fig:fig2}}

\placefigure{Fig. \ref{fig:f12}}

\subsubsection{\label{sec:cloudinfo}Components HV0--HV7}

These features are isolated knots or diffuse edges of bright clouds
in the FOC {[}\ion{O}{3}{]} image,
and are labelled
0--7 in Fig.~\ref{fig:fig1}.
Their line profiles in the left columns of Fig.~\ref{fig:f12} show:

\begin{enumerate}
\item A component whose flux centroid redshifts from systemic velocity
at knot 0 to 600 km~s\( ^{-1} \) at knot 7. It has
two FWHM values: in knots 0--3 it is 300--400 km s\( ^{-1} \),
while in knots 4--7 it is 900 km s\( ^{-1} \).  The descriptions that follow
omit this component.
\item An extended blue wing out to --3200 km~s\( ^{-1} \), whose behavior
in the different gas complexes we now detail. 
We refer to these as ``high velocity" features HV0--HV7.
\end{enumerate}

\paragraph{Complex 0}

is sampled by slits 1 \& 2. In the FOC image, it is the diffuse, SE, boundary
of the brighter, central NLR.
Flux peaks near systemic velocity in the SW part of this complex
and then extends to --2700
km~s\( ^{-1} \) as the distance from the nucleus increases by
\( 0\farcs 2 \) (14 pc).

\paragraph{Complex 1,}

centered \( 0\farcs 5 \) (35 pc) NE of complex 0, is a remarkable
cluster of roughly a dozen unresolved knots in the FOC image. The knots
subdivide into a comparable number of bright, blueshifted peaks in each of slits
1 to 3, and we can usually match a knot to a velocity component because the
knots are separated along the slit. The spectrum of the knot in slit 1 shows
a diagonal feature, which is resolved into a series of spatially resolved
steps in space-velocity that each span \( \sim  \)400 km~s\( ^{-1} \),
indicating an increase in gas velocity with position
along the slit from $\sim -400$ km~s$^{-1}$ to $\sim -2500$ km~s$^{-1}$
over 0\farcs5 (35 pc).
In slits 2 and 3, two increases with nuclear distance occur: from
--250 to --1250 km~s\( ^{-1} \) then --1200 to --3200 km~s\( ^{-1} \)
over the \( 0\farcs 4 \) (28 pc) extent of the complex.
Despite multiple discrete spatial components, the overall 
emission-line profile is smooth.

\paragraph{Knot 2}

appears only in slit 4, where it is distributed between --1800 and
--300 km~s\( ^{-1} \). It is elongated slightly N--S in the FOC image.

\paragraph{Knot 3}

is a spatially diffuse knot in slit 4, with gas between --100 and
--2400 km~s\( ^{-1} \). The trend along slit 4, through knots 2 and
3 and then farther NE, is that of an incomplete {}``Doppler ellipsoid''
centered at --600 km~s\( ^{-1} \) and extending around that velocity
by \( \pm 1500 \) km~s\( ^{-1} \).  In the FOC image (Fig.\ 1) it appears as
an elliptical ring of diameter 0\farcs55 centered 2\farcs2 from S1
(50 pc) delineated by a half-dozen
faint wisps (see pink broken ellipse in Fig. \ref{fig:fig1}). This
apparently disintegrating structure therefore has a dynamical age
of \( \sim 1.3\times 10^{4} \) yrs.

\paragraph{Complex 4}

is included in slit 5,
has 3 or 4 unresolved knots in the FOC image, and lies
along the edge of complex
G. Velocity features are evident in slit 5, and
extend in two adjacent bands, from --200 to --2900 km~s\( ^{-1} \)
and from --1200 (faint) to $<$ --3000 km~s\( ^{-1} \).

\paragraph{Complex 5}

forms the diffuse, NE boundary of clouds G and H.
We may just resolve it in slit 6 into a linear
gradient from --200 to --2300 km~s\( ^{-1} \) as nuclear distance
increases.

\paragraph{Knots 6 \& 7}

lie outside the area mapped by our six parallel spectra, but are intersected
by the archival observation with 
slit along P.A. 10\arcdeg. These knots resemble others
in the FOC image, with knot 6 extending to --1800 km~s\( ^{-1}\) and
knot 7 extending to --1600 km~s\( ^{-1} \).

\subsubsection{Clouds}

Our spectra in the right-hand columns in Fig. \ref{fig:f12}
include all the large clouds first delineated by \cite{ev91}.
Several clouds were spanned
by CKa with their single L-grating slit, so the local continuum properties
that those authors established are also mentioned in the following remarks.

\paragraph{A,}

included in slit 3, has a small peak centered at 500 km~s\( ^{-1} \),
in addition to weaker emission that spans --1900 to +1000 km~s\( ^{-1} \).

\paragraph{B,}

included in slit 4, extends from \( < \) --2200 to \( > \)+2500
km~s\( ^{-1} \) (the velocity extent is uncertain because
of the difficult continuum subtraction in our narrow spectral bandpass).
The emission extends \( \sim 0\farcs 2 \) along the slit. 
The associated continuum is
the {}``hotspot'' discussed in detail by CKa that is probably
scattered nuclear light.

\paragraph{C,}

included in slit 5, extends from --1400 to +2000 km~s\( ^{-1} \).
CKa found that most of its continuum is scattered nuclear light.
\cite{bi98}
propose that this cloud is being agitated by knot C in the radio jet,
which \cite{ro01} argue is a shock standing in the jet, based on
its low proper motion in multi-epoch radio VLBI images.

\paragraph{D}

is also included in slit 5. A faint unresolved component reaches --1800
km~s\( ^{-1} \) and attaches to brighter, spatially extended emission
that spans --500 to +1800 km~s\( ^{-1} \). CKa found that half of its continuum
is scattered nuclear light.

\paragraph{E,}

included in slit 6, spans --1500 to +1300 km s\( ^{-1} \), but
is brightest between --500 and +100 km s\( ^{-1} \).

\paragraph{F,}

included in slit 3, has two separate plumes. That 
nearer to the nucleus
extends from systemic velocity to --2700 km~s\( ^{-1} \), that
farther extends from --400 to +2000 km s\( ^{-1} \).

\paragraph{G,}

included in slit 5, starts abruptly at --200 km~s\( ^{-1} \) and extends
to +1400 km~s$^{-1}$. 
It emits strong {[}\ion{Fe}{7}{]} (PFBW).
At lower spectral resolution, PFBW and
KC1 both found a very large flux ratio {[}\ion{O}{3}{]}/H\( \beta =22 \)
near peak emission. However, our high-resolution spectra isolates
this component more effectively, reducing the summed flux ratio to $\sim$17.

\paragraph{H}

is sampled by slit 6.
Its profile resembles that of cloud G, but is
skewed to brighter emission at blueshifts between --500 and --1300 km~s\( ^{-1} \)
vs. --250 to --900 km s\( ^{-1} \) for G. CKa found that 
half of its continuum is scattered nuclear light.

\paragraph{J \& K}

are covered by slits 4+5 and 6, respectively.  They lie on straight lines drawn
back to the nucleus from, respectively,
clouds G and H. \cite{Ax97} discovered that these features are double
peaked, and we find that they are centered first at +500 km~s\( ^{-1} \)
then at --700 km~s\( ^{-1} \) as distance from the nucleus increases.

\paragraph{I}

is the brightest line-emitting part of gas associated with the SW radio lobe.
Line profiles are double-peaked, with the stronger
peak redshifted (near +700 km~s$^{-1}$) and sharp in slits 1-3, but more 
diffuse in slits
4--6. As the nuclear distance increases by 1\arcsec, velocity centroids
shift from +250 to +800 km~s\( ^{-1} \) in slit 5 and from systemic
to --500 km~s\( ^{-1} \) in slit 6. 

\subsubsection{Clouds Associated with the NE Radio Lobe}

Our slits sample many filaments across the base of this extended bow
shock \citep{Wi87} in the interval 2--4\arcsec\ from the nucleus. Some
filaments beyond clouds G and H and a few fainter knots beyond the
top of Fig.~\ref{fig:fig1} are blueshifted, but beginning at 2\farcs6
radius most of what appears near the top of the FOC image as a coherent
V-shaped feature is redshifted in slits 1--6, confirming the 
findings of PFBW. Between \( 2\farcs 9 \) and 4\arcsec\ radius,
it is striking that individual components are resolved along the slit
yet all range in velocity from systemic to +1200 km~s\( ^{-1} \).
There is a narrow feature near systemic velocity and only faint, diffuse
emission to the blue.
From the {[}\ion{S}{2}{]} doublet in this region, PFBW derived gas
densities near systemic velocity of \( \sim 400 \) cm\( ^{-3} \) if
\( T_{e}=10^{4} \) K. 

\placetable{Table 2}

\subsection{Physical Properties}

We have used the line profiles displayed in Fig.~3 to constrain the ionized
gas mass, kinetic energy, and momentum of each velocity component.
A mask was used to isolate the component in the higher signal/noise
ratio {[}\ion{O}{3}{]} composite profile. This mask was then applied to
the same velocity range in the dereddened H$\beta$ profile.  The resulting
flux has been converted into an ionized gas mass,
assuming case-B recombination conditions (for which KC1 found evidence
in the red-wing component NE of the nucleus) and a pure H plasma.
This conversion depends on the average electron density.
In Table 2 we have therefore scaled some quantities by density \( n_{e,4}=10^{4} \)
cm\( ^{-3} \), which \cite{do97} found was appropriate for the NLR.
Undoubtedly, a large range of gas densities are present in
each emitting feature. 

The high-velocity features HV0--HV7
emit a few percent of the total NLR H\( \beta  \) flux.
In Table 2 we also tabulate properties of clouds A--H that were 
identified by \cite{ev91}, 
and introduce clouds J and K (cloud I is rather ill-defined,
in the SW radio lobe).

\section{Discussion}

While the clouds might have been as blended at \emph{HST} resolution
as they are from the ground, in \S3 we showed that our deep spectra
have resolved spatially the NLR. We have also resolved much of it 
kinematically, although not
down to the thermal width (17 km~s\( ^{-1} \) in \( 10^{4} \) K
gas) in part because of inadequate spectral resolution
and in part because several components are still projected along each line
of sight and may have turbulent substructure. 

To track the energy flow and to isolate possible shock-excited structures,
we must now constrain the internal properties and space velocities
of the NLR clouds. 

\subsection{\label{sec:photo}Photoionization of the NLR Clouds}

The NLR clouds are bright features 
embedded in the large-scale ionization cones and
are distributed over several arc-seconds, so have been studied extensively
from space and, in aggregate, from the ground. FOS spectroscopy lacked
spatial resolution to separate clouds from their surroundings.
Subsequently, KC obtained a STIS spectrum along one cut
through the NLR using low-resolution gratings that spans
\( \lambda  \)\( \lambda  \)115
-- 1027 nm. Along the slit, they analyzed continuum fluxes (CKa), line
fluxes (KC1), and gas kinematics (CKb). In KC1 they reproduced the
observed flux ratios with AGN photoionization models that require
considerable absorption along sight-lines from the nucleus that intersect
the galaxy gas disk. They confirmed a correlation between ionization
potential and line blueshift from galaxy systemic velocity that \cite{mo96}
had discovered in their ground-based IR spectra. 

KC1 sampled several clouds NE of the nucleus, so their photoionization
models also address the new features that we have mapped. 
Their models combine
matter- and ionization-bounded clouds, the former arising plausibly
from low-density, photoevaporated envelopes of the latter \citep{bi96}.
Matter-bounded
clouds may filter AGN photons that impinge on the ionization-bounded
cores, so internal dust would play a critical role in setting their
emission-line flux ratios \citep{DoGr01,gr01}. 

The models of KC1 required that the blue-wing clouds be optically
thin above the Lyman 
limit, and need another photon source between 1\farcs4-1\farcs8
to explain strong NUV lines and a larger \ion{He}{2}\( \lambda  \)4686/H\( \beta  \)
ratio along 202\arcdeg\ P.A. 
CKb suggested that clouds are decelerating here, forming
a \( \sim  \)1000 km~s\( ^{-1} \) shock that would brighten
high-ionization lines as observed. They lacked the sensitivity
and spatial coverage of \cite{mo96} to coronal-line emission, so unlike those 
authors found no such emission beyond cloud B and the UV continuum hotspot.
In CKa they established that the scatterer is tenuous and coincides
with the deceleration region. They found reasonable fits to blue-wing
line fluxes if 85--90\% of the flux comes from tenuous gas with 
\( (\log \mathcal{U}, \)\( n_{H},\log N_{H})=(-1.45\pm 0.15,3\times 10^{3}\, \rm{cm^{-3}},21.18\pm 0.18\, \rm{cm^{-2}}) \) 
and the rest from denser clouds with \( (\log \mathcal{U} \)\( ,n_{H},\log N_{H})=(-2.9,2\times 10^{4}\, \rm{cm^{-3}},20.7\pm 0.1\, \rm{cm^{-2}}) \),
both components needing dust fractions of 10--25\% (larger values closer
to nucleus) to suppress Ly\( \alpha  \) and C IV to their observed dereddened
fluxes.

KC1 found that the red-wing clouds have the higher dust fraction (25\%)
and are exposed to a lower ionizing flux (especially above the \ion{He}{2}
Lyman limit), indicating absorption of AGN photons on the way
to the red wing gas, and thence by the tenuous blue-wing component
along our line of sight. \cite{mo96} find IR coronal lines across
the inner 4\arcsec\ diameter that emit out to 2000 km~s\( ^{-1} \)
blueshift from systemic velocity, and which can arise either in gas
photoionized by a hard UV continuum or in a hot (\( \sim 10^{6} \)
K) collisionally ionized (perhaps shocked) plasma. \citet{ol01} note that
the {[}\ion{Fe}{2}{]} emission
is unlikely to come from shocks because they find that the flux ratio
{[}\ion{Fe}{2}{]}/{[}\ion{P}{2}{]}$\sim1$, 5\% of the value seen
in supernova remnants. 

X-ray results also support photoionization NE of the nucleus, beyond
radius $r_i$. First,
\cite{Yo01} show that filaments in the \emph{Chandra} ACIS image correlate well
with those in [\ion{O}{3}]$\lambda$5007 and H$\alpha$+[\ion{N}{2}] images. 
Second, the \emph{Newton/XMM} RGS spectrum reported
by \cite{Pa00} \cite[also][]{be01} show that gas in this region is
photoionized, based on the detection of narrow radiative recombination
continua, line flux ratios within the helium-like triplets, and the
weakness of Fe L-shell compared to K-shell emissions.

\subsection{\label{sec:velocities}Deprojected Velocity Field}

\begin{figure}
\epsscale{0.6}
\plotone{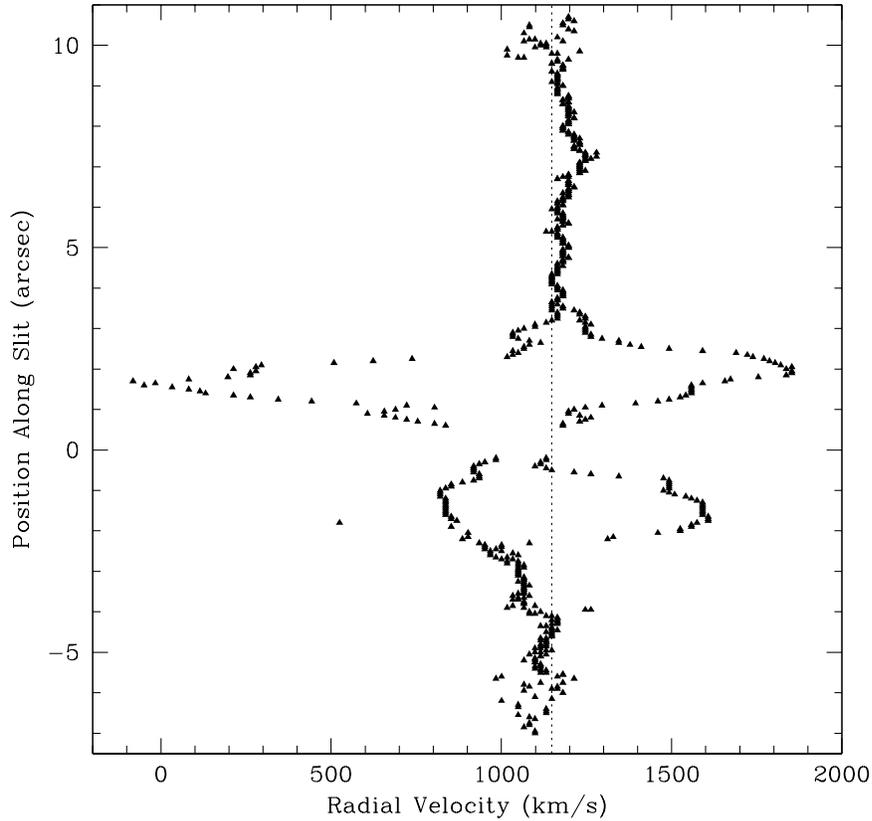}
\caption{\label{fig:velocities}Velocities of the flux centroids of the 
two main emission-line
components in each continuum-subtracted line profile as a function of distance
from the nucleus, obtained from an
archival M-grating spectrum (slit P.A. = 10\arcdeg).
NE is at top and SW is at bottom. The vertical dotted
line shows the systemic velocity of the galaxy. The velocity field
is similar to that shown in Fig.\ \ref{fig:fig2}
of CKb that was extracted
along a slightly different P.A. The high-velocity, blueshifted motions 
between 1\farcs5--2\farcs5
radius NE of the nucleus result from knots 6 and 7.}
\end{figure}

Ground-based Fabry-Perot (CBT) and integral-field
spectra (PFBW) have constrained the bulk motion and space distribution
of the NLR clouds. CBT found an overall distribution of
clouds in a thick bicone with maximum
half opening angle \( \sim 41\arcdeg  \) and whose 
axis is inclined \( \sim 85\arcdeg  \) to our line of sight; PFBW showed
that gas coincident with
the NE radio lobe beyond 2\farcs5 radius is predominantly redshifted.
CKb also posited a biconical flow
to match the velocity centroids along their single STIS slit,
requiring outflow velocities that reach 1300 km~s\( ^{-1} \) 
at $r_i\sim1\farcs7$ NE of the nucleus and then decrease beyond.
Limited spatial coverage of a structure whose symmetry axis lies near
the plane of the sky meant that their kinematical data are consistent with
either an outflow that is radial from the nucleus or one that is
roughly perpendicular to the jet axis. 
The archival M-grating spectrum (Fig.\ \ref{fig:velocities})
shows that knots HV6 and 7 are comparable to the largest blueshifts in
the model geometry of CKb.

\placefigure{\ref{fig:velocities}}

\begin{figure}
\epsscale{1}
\plotone{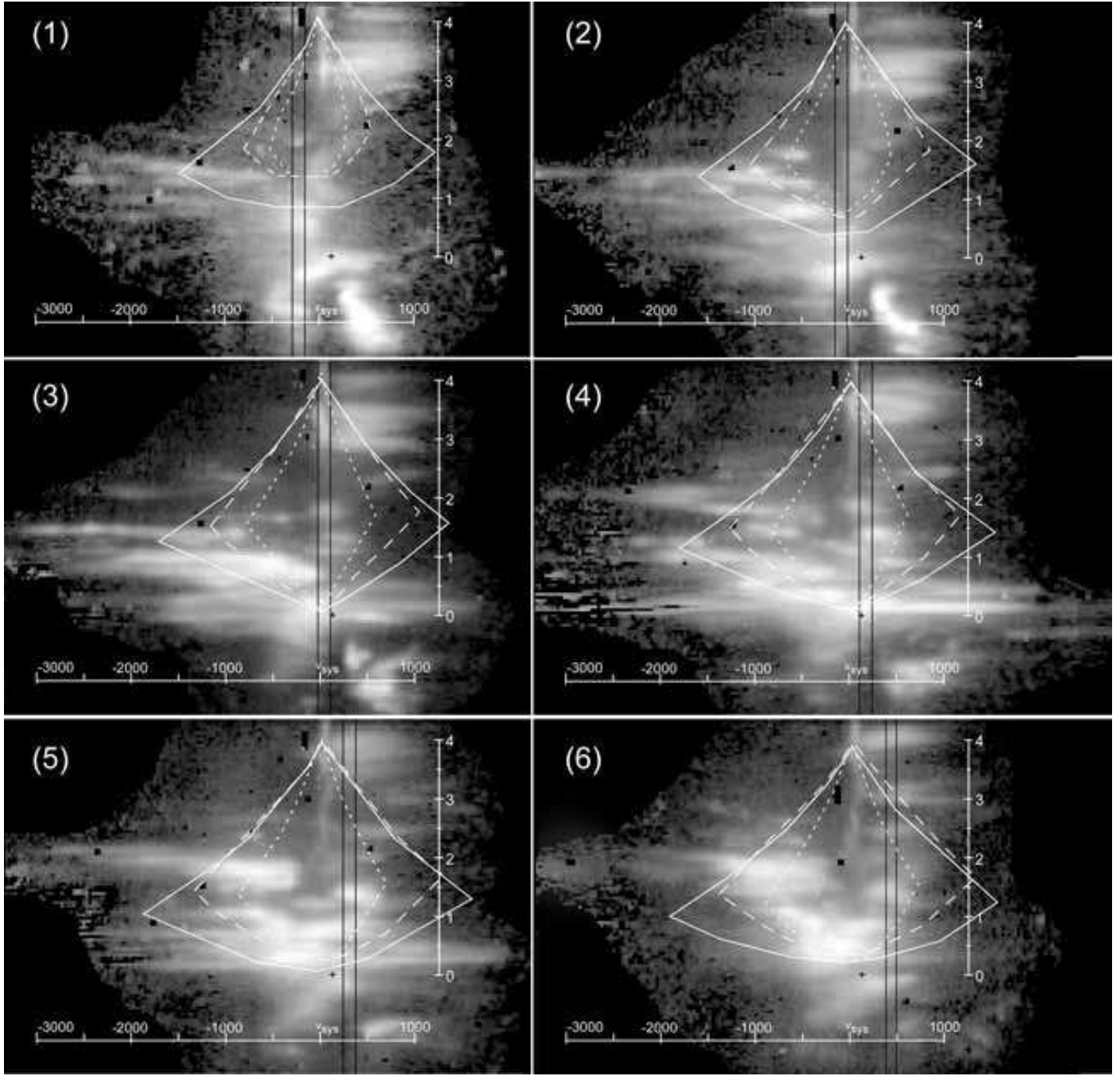}
\caption{\label{fig:fig5}Same display format and region as for panels 1-6 in
Fig.\ \ref{fig:fig2}, but now with contours to show
the space-velocity projections of the outer cones in the three
bi-conical outflow models described in the text: 
(solid lines) increasing
the maximum outflow velocity from the nucleus to 2500 km~s\( ^{-1} \)
at $r_i$,
(long dashed lines) having the gas expand cylindrically from the
jet axis at close to the observed velocity, (short dashed lines)
outflow model of CKb. At best these models match only the blue wing of the data
within 1\farcs5 NE of the nucleus. The red wing is not fit at all.}
\end{figure}

We see abrupt jumps in velocity as the STIS slits cross the clouds and
knots evident in the FOC narrow-band image. 
In Fig.\ \ref{fig:fig5}, we compare our line profiles with the projections
of several models of radial and cylindrical outflow.
By increasing the half angle of the NE bicone
from \( \sim  \)40 to 50\arcdeg\ within the radius \( r_{i} \) of the
CKb model where outflow velocities begin to decrease, we better match the
trend in the centroids of radial velocities of the blueshifted features
in our spectra. The velocity
limits of our two models are shown in the Figure, and correspond to: 
(solid line) increasing
the maximum outflow to at least 2500 km~s\( ^{-1} \) at $r_i$, and
(long dashed line) having the gas expand from the
jet axis with all of its motion along our line of sight.
We also plot the outflow model of CKb in short dashed lines.
The gas becomes increasingly blueshifted (on average) out to radius 2\arcsec,
implying its acceleration on that scale. 
However, the observed gas does not fill all of the NE bicone,
because redshifted emission is observed at much lower velocities than
predicted by all of the outflow models until the NE radio lobe is reached.
Emission then jumps to a range that extends from zero to $>1000$ km~s$^{-1}$
redshift (relative to systemic velocity)
and persists from 3\arcsec\ to the middle of the NE radio lobe
(which occurs at the top of the panels in Fig.\ 2); the emitting features
are spatially compact.
KC1 noted
that the emission-line flux ratios of the red wing gas are consistent with
irradiation by a heavily absorbed component. If these clouds are accelerated
radiatively, they may not attain velocities as high as those that were
irradiated directly.

\placefigure{\ref{fig:fig5}}

\begin{figure}
\epsscale{0.6}
\plotone{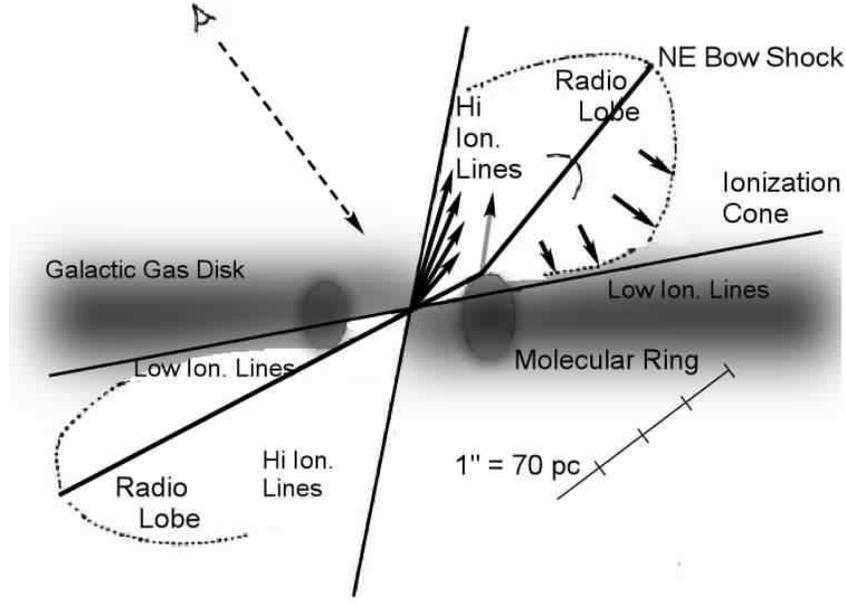}
\caption{\label{fig:fig6}Kinematic model of the NLR refined from that proposed
by \cite{te01}.  To represent patterns seen in our spectral maps, we have 
(i) added radial expansion from the nucleus, the velocity
of that outflow appearing to increase with its angle of travel above the disk;
(ii) removed blueshifted gas associated with the expanding radio 
lobe because our data show only redshifted gas that is being 
pushed into the galaxy disk; (iii) added the gray arrow to represent
possible ablata from the molecular ring (which may be a highly warped 
outer extension of the putative torus obscuring our view of the AGN).}
\end{figure}

\placefigure{\ref{fig:fig6}}

As suggested by \cite{te01} based on infrared [\ion{Fe}{2}]
spectra and PFBW based on spatially deconvolved optical spectra,
the red wing gas beyond 2\farcs5 radius is apparently 
being pushed into the dense galactic
disk by the lateral expansion of the NE radio lobe.
Our favored scenario is shown in Fig.\ \ref{fig:fig6}.
The jet is inclined to the disk.  Eventually,
at radius 2\arcsec, it bursts out of the disk and inflates the radio lobe. The
lobe expands laterally in all directions, but is only visible
where it pushes into the disk and creates the zero to redshifted emission.
The large blueshifted extensions on the high-velocity
clouds suggest that the clouds are
moving slower than the HV knots. The HV flow must have been accelerated
relative to the clouds or reoriented closer to our line of
sight by a combination of bouyancy, radiative acceleration, and lobe/cloud 
collisions \citep{ta92}.  Depending upon their dynamics, the larger
clouds may lag the knots because of larger drag forces, as KC1 have proposed
for the red wing component.

\subsection{Acceleration of Clouds \& Knots}

\subsubsection{\label{sec:jet}Jet Interactions with Clouds C \& D}

Using the long-slit mode of the FOC, \cite{Ax97} found that line profiles
are double-peaked and have highest gaseous excitation across the radio
jet near clouds G and what we call K. On this basis they argued that
hot shocked gas in the jet cocoon expands clouds from the jet axis.
Nearby, \cite{kr98} found evidence from emission-line ratios in an
FOS spectrum for cosmic-ray heating of the emission-line gas. The
Merlin and VLA images of \citet[G96 hereafter]{Ga96}, when registered
using the astrometry of \cite{ca97} (see red contours in Fig.\ 1),
place NLR clouds B, C, D, and F adjacent to the radio jet; in particular,
clouds B and C straddle a stationary radio knot \citep{ro01}
where the jet bends \( \sim 23 \)\arcdeg. \cite{bi98} have modeled
this complex as a jet-cloud interaction. What constraints do our data
impose? 

KC1 found that cloud C is photoionized. However, decomposing
its profile in Fig. \ref{fig:f12} into Gaussians isolates a component
centered \( \sim 300 \)
km s\( ^{-1} \) blueward of systemic and a wing component centered
300 km s\( ^{-1} \) to the red, both with dispersions \( \sim 500 \)
km s\( ^{-1} \).  The blue component has flux
ratio {[}\ion{O}{3}{]$\lambda5007$}/H\( \beta \sim 16 \) \emph{vs.}
\( \sim 8-10 \)
for the red component. 
While the considerable line widths are consistent with shocks,
the flux ratios are
consistent with photoionization by the AGN with the red wing gas more 
absorbed. 
Thus, deflection of the jet at cloud C is not clearly connected
to the excitation of the cloud.
Cloud E, which is farther
from the jet than cloud C, shows a similar blue feature but not one
in its red wing.

The NE component of the nuclear radio triple is 0\farcs75 NE of
knot S1 (the nucleus).
G96 note that this jet component, like the radio knot near cloud D,
has a flat synchrotron spectrum and extended emission that is misaligned with
the jet.
However, the jet does not bend at the NE component nor is there maser emission,
arguing against a shock front. It is therefore interesting that the
emission-line profiles of cloud D are double peaked. The blue peak
has flux ratio {[}\ion{O}{3}{]}/H\( \beta \sim 15 \), but now the
red peak ratio decreases monotonically through the profile from 17
to $<3$ over a 1000 km s\( ^{-1} \) range. This decline implies that
gas density increases with the velocity deviation from galaxy systemic,
not what would be expected if most of the cloud mass is at velocities
close to rest in the galaxy.
G96 note that a cloud interacting with the
jet must be sufficiently massive to avoid acceleration, implying
\begin{eqnarray*}
M_{cloud\, B}\geq 40\, M_{\bigodot }\left( \frac{r_{j}}{7.5\, pc}\right) ^{2}\left( \frac{\rho _{j}}{5\times 10^{-27}\, g\, cm^{-3}}\right) \times & &
\end{eqnarray*}
\begin{eqnarray*}
\left( \frac{v_{j}}{15,000\, km\, s^{-1}}\right)
\left( \frac{t_{lobe}}{3.5\times 10^{5}\, yr}\right). &  & 
\end{eqnarray*}
Here jet density and velocity are normalized to the values appropriate
for the lighter jet model of \cite{Wi87} for the NE radio lobe, $t_{lobe}$
is the age of the NE radio lobe from spectral aging as estimated by G96, and
the jet radius \( r_{j} \) was measured by \cite{ga94}. If the average NLR
cloud density is \( \sim  \)150 cm\( ^{-3} \), the
entries in Table 2 show that clouds are massive enough to remain adjacent
to the deflecting jet. This value is consistent with the mean densities
of Galactic molecular clouds \citep{so87}.

Elsewhere, radio and line emission are uncorrelated. G96
note that this is expected if the jet has swept all gas from the region.
The high-velocity knots might then be ablated from molecular
clouds in the galaxy disk that have rotated into the jet. The clouds
accelerated would all be blueshifted because they would move predominantly
away from and above the galaxy disk. Similar clumps below the jet
would sink into the denser disk where they would be subject to stronger
drag forces and hence decelerated more effectively (consistent with the
gas distribution in the spectral maps of PFBW).

\subsubsection{Shocks Near the X-ray Subpeak?}\label{vshape}

Three characteristics of the \emph{XMM/Newton}
X-ray spectrum of NGC 1068 within 20\arcsec\ of the nucleus indicate that
the gas is predominantly photoionized:
(i) narrow radiative recombination continua; (ii) in the
He-like triplets the forbidden lines are stronger than the resonance
lines; and (iii) Fe L emission is weak compared to the K shell emission
of the abundant elements \citep{ki01,be01}. 
However,
subtraction of the predictions of a photoionization model from the spectrum
shows residuals that may indicate a collisionally excited component
or other effects \citep{Pa00}.

Moreover, near the V-shaped line-emitting feature whose apex is 
3\arcsec\ NE of the nucleus (at top of Fig.\ 1), the
\( 4-26 \) \AA\ spectrum from the \emph{Chandra} HETGS \citep{og01} shows
that resonance lines are brighter than forbidden lines, implying either
a collisionally ionized component or strong resonant scattering.
\cite{og01} show that the collisional component lies at the X-ray
subpeak reported by \cite{Yo01}.
The X-ray subpeak may represent gas in the galactic disk on the far side of the
NE lobe and being compressed by it (cf.\ PFBW).
As discussed in \S\ref{sec:photo}, photoionization models of the
UV and optical STIS spectra of KC1 also support this additional power
source near the subpeak. 

\subsubsection{\label{sec:shocks}Shocks Near Clouds G-K?}

Two lines of evidence argue for shocks near clouds G-K:

\begin{enumerate}
\item Clouds G and K are close to the ``NE radio subpeak", located
2\farcs1 NE of the obscured nucleus \citep{Wi87}, suggesting dissipation of
kinetic energy and possibly cloud acceleration by the radio jet.
\item Nearby we find the Doppler ellipsoid described in \S\ref{sec:cloudinfo},
whose expansion velocity of 1500 km~s\( ^{-1} \) will generate a post-shock
temperature of \( \sim 3\times 10^{7} \) K if the ambient medium
is relatively stationary, fully ionized, and has 10\% He abundance. 
\end{enumerate}

To estimate the mechanical luminosity of the Doppler ellipsoid, 
we assume that energy is injected at constant rate, that the shell shock
is strong so that we can ignore the thermal pressure of the ambient
gas, and that the bubble is within the scale height of the galaxy
disk so that the ambient ISM density is constant. Then in the energy conserving
phase of the bubble interior, the bubble radius and expansion velocity
are related: \( R=1.67vt. \) By observation, \( t=0.012 \)
Myr. Mechanical luminosity \( L \) follows from \( R(t)=31(L_{36}/n_{0})^{0.2}t_{6}^{0.6} \)
pc = 100 pc with \( t_{6}=0.012 \) (where \( L_{36} \) has units \( 10^{36} \)
ergs s\( ^{-1} \) and \( n_{0} \) is the density of the ambient
ISM in cm\( ^{-3} \)), so \( L=5.2\times 10^{7}n_{0} \) L\( _{\odot } \).
Overall, the filaments that delineate the bubble and maybe its associated
ISM shock emit $<$2\% of the NLR {[}\ion{O}{3}{]} and H\( \beta  \)
fluxes.

\subsubsection{\label{sec:accel}Radiative Acceleration of Dusty Clouds}

Dust absorption dominates opacity in a photoionized plasma like this
NLR when ionization parameter \( \mathcal{U} >\alpha (T_{e})/c\kappa  \),
where \( c \) is lightspeed, \( \kappa \sim 10^{-21} \) cm$^2$ atom$^{-1}$
is the opacity
assuming the standard Galactic reddening curve, and \( \alpha (T_{e})\sim 2\times 10^{-13} \)
is the recombination coefficient. The dust will compete successfully
for ionizing photons in the highly ionized zone, suppressing line flux.
A strong pressure gradient will also be
set up so that the plasma density will be much enhanced by the time
lower ionization species emit. In consequence, a lower
\( \mathcal{U} \sim 0.007 \)
will provide an apparent model fit to the spectrum, close to the value
preferred by KC for their denser component.
This ionization parameter
is closely related to the value of \( \mathcal{U} \) at which radiation
pressure starts to dominate either the pressure gradient, or the dynamical
acceleration of the ionized plasma. Thus, the flow of dusty clouds
falling toward the nucleus is intrinsically unstable above the dust
sublimation point, with clouds being driven into outflow in the direction
of slowest accretion (i.e.\ within the ionization cones). 
\cite{DoGr01} detail this scenario.  In hydrodynamical
simulations of cloud acceleration \citep{sc95}, tenuous gas streams
ablate from surfaces of massive clouds, which may be the source of the
high-velocity features we observe. 

In this NLR, the flux in the ionizing continuum needed to maintain
the observed line luminosity (\( \sim 3\times 10^{44} \) ergs~s\( ^{-1} \),
KC1) is comparable to the FIR luminosity 1.5\( \times 10^{11} \)
L\( _{\odot } \). Dusty clouds are accelerated radiatively by \( \sim 10^{-5} \)
cm~s\( ^{-2} \), so can reach their observed mean velocities of
\( \sim 500 \) km~s\( ^{-1} \) over 35 pc. In Fig.~3,
the cloud line profiles indeed show less velocity structure than
do the knots.

This scenario ignores the role of the radio jet in agitating the NLR clouds
\citep{bi98}. But, as discussed earlier, its influence
may be localized to line broadening and ablation in specific jet/cloud
and lobe/cloud interactions, for example between B and C \citep{Ga96} and
G and H \citep{Ax97}. While
the region of collisional line emission coincides with the base of
the NE radio lobe, numerical simulations of a breakout lobe do not
show (D.\ Balsara, private communication) a strong backflow vortex
at its base that might form high-velocity shocks. In any event, many
of the high-velocity knots have similar line profiles, and are found
\( >40 \) pc from the brighter parts of the jet.

\section{\label{sec:absorbers}Are the High-Velocity Knots {}``Associated
Absorbers'' Seen in Emission?}

If the HV emission knots in NGC 1068 were instead viewed against the
nuclear continuum, they would resemble in their kinematics the
Associated Absorption Line (AAL) systems seen in a few percent
of quasars \citep[for example]{ha97}.
KC1 argue from their spectral models that the putative UV absorber in the NE 
quadrant must block less than half of the nuclear light.
If the blueshifted knots have the same extent
along the line of sight as we see on the sky (assumed here to be just
resolved), they would cover at most 2\%
of the sky at the nucleus, consistent with values derived for some
AAL absorbers. 

Such absorbers are usually thought to lie much closer to the
source of hard photons than the 35--160 pc radial distance of the
clumps in NGC 1068.  However, recent photoionization
models of high-resolution quasar spectra are establishing that AAL
absorbers lie at diverse distances from the AGN, including a range
comparable to those we see. For example, \cite{dk01} established
that some absorbers in a BALQSO must be \( \sim  \)700 pc from that
nucleus and have kinematic structure similar to the HV components
in NGC 1068. High-velocity features are also visible on similar
scales in STIS spectra of NGC 4151 \citep{hu99}, although
at more modest velocities (1400 km s\( ^{-1} \)) than we see, and
are not clearly connected to massive clouds as in NGC 1068.

\section{Future Work to Constrain Filament Properties}

The brighter blueshifted knots in NGC 1068 could be mapped
in higher ionization lines with UV spectra to establish their
optical depths and ionization structure. But this study would be arduous
with STIS because of their relative faintness and large reddening.
Unfortunately,
the complex background emission makes the knots unsuitable targets
for the high-efficiency Cosmic Origins Spectrograph (COS) to
replace STIS on \emph{HST}. 

Now that specific targets are known up to 2\arcsec\ from the nucleus,
practical spectroscopy is possible with optical and near-IR
lines using integral-field spectrometers
that are appearing on 8--10m telescopes. An interesting constraint
follows if such spectra find evidence of the H~I/II recombination
front present in optically thick clouds \citep[for example]{sh95,vi89}. Constraining
the column will better constrain photoionization models to predict
the full mass --- hence KE, momentum, and ultimately the origin --- of each
knot and cloud.

\section{Summary}

Our STIS spectral maps have resolved the spatio-kinematic structure of
the NLR in NGC 1068, and reveal that many compact knots in the FOC
image have large, exclusively blue-shifted velocities, ranging up to
$\sim3200$ km~s$^{-1}$ from systemic. If these knots
are optically thick in the UV continuum, they are good candidates
for ``associated absorber" clouds present in other AGN. Lying 70--150 pc from
the nucleus, several form a broken spatio-kinematic ring of diameter
35 pc that is expanding at up to 1500 km~s\( ^{-1} \).
The ring is adjacent to the more massive but slower moving NLR clouds
G--K, so may originate from ablata streams from disintegrating molecular
clouds that are being photoionized and accelerated radiatively by
the AGN or mechanically by a knot in the radio jet.

By resolving this NLR, we have demonstrated that ground-based
integral-field spectrometers will be able to constrain the properties of the
emitting gas in NGC 1068 by
using classical plasma diagnostics in the optical
and near-IR. More fundamentally, we may have a convenient view onto gas 
associated with
an important component of AGN that is hard to study elsewhere because
it is usually seen in absorption.

\acknowledgements{}

This work is based on observations made with the NASA \emph{Hubble Space 
Telescope},
obtained at the Space Telescope Science Institute, which is operated by the
Association of Universities for Research in Astronomy, Inc., under NASA
contract NAS5-26555. These observations are associated with proposal ID
GO-7353.
We thank Alessandro Capetti for the \emph{HST} FOC mosaic of NGC 1068, and NASA
for support through grants NAG 81027 and GO-7353.
GC thanks the Research School of Astronomy and Astrophysics in the 
Institute for Advanced Studies of the Australian National University, and its 
Director Prof.\
J. Mould, for financial support and a stimulating work environment to begin
this paper.

\onecolumn

\begin{deluxetable}{lcccccc}
\tablecaption{STIS Long-Slit Observations of NGC 1068}
\tablewidth{0pc}
\tablehead{ & & & \colhead{Dispersion} & \colhead{Spectral} & \colhead{Exposure} & \colhead{Slit P.A.} \\
\colhead{Data Set} & \colhead{Date} & \colhead{Grating} & \colhead{(\AA\ pixel$^{-1}$)} & \colhead{Range} & \colhead{(s)} & \colhead{(deg.)} }
\startdata
o56502010 & 1999 Oct 2 & G430M & 0.6 & 4818-5104 & 2585 & 38 \\
o56502020 & 1999 Oct 2 &G430M & 0.6 & 4818-5104 & 2775 & 38 \\
o56502030 & 1999 Oct 2 &G430M & 0.6 & 4818-5104 & 2775 & 38 \\
o56502040 & 1999 Oct 2 &G430M & 0.6 & 4818-5104 & 2775 & 38 \\
o56502050 & 1999 Oct 2 &G430M & 0.6 & 4818-5104 & 2772 & 38 \\
o56503010 & 2000 Sept 22 &G430M & 0.6 & 4818-5104 & 2294 & 38 \\
o56503020 & 2000 Sept 22 &G430M & 0.6 & 4818-5104 & 2853 & 38 \\
o56503030 & 2000 Sept 22 & G230L & 1.6 & 1568-3184 & 2376 & 38 \\
o56503040 & 2000 Sept 22 &G230L & 1.6 & 1568-3184 & 2376 & 38 \\
o56503050 & 2000 Sept 22 &G230L & 1.6 & 1568-3184 & 2376 & 38 \\
o56503060 & 2000 Sept 22 &G230L & 1.6 & 1568-3184 & 2376 & 38 \\
o56501010 & 1999 Sept 22 & G140L & 0.6 & 1140-1730 & 1425 & 38 \\
o56501020 & 1999 Sept 22 & G140L & 0.6 & 1140-1730 & 1425 & 38 \\
o56501030 & 1999 Sept 22 & G140L & 0.6 & 1140-1730 & 1425 & 38 \\
o5lj01050\tablenotemark{a} & 2000 Jan 14 & G430M & 0.3 & 4818-5104 & \phn800 & 10 \\
o5lj01060\tablenotemark{a} & 2000 Jan 14 & G430M & 0.3 & 4818-5104 & \phn846 & 10 \\
o5lj01070\tablenotemark{a} & 2000 Jan 14 & G430M & 0.3 & 4818-5104 & \phn960 & 10 \\
o5lj01080\tablenotemark{a} & 2000 Jan 14 & G430L & 0.6 & 2900-5700 & \phn720 & 10 \\
o5lj01090\tablenotemark{a} & 2000 Jan 14 & G430L & 0.6 & 2900-5700 & \phn720 & 10 \\
o5lj010a0\tablenotemark{a} & 2000 Jan 14 & G430L & 0.6 & 2900-5700 & \phn821 & 10 \\
\enddata
\tablenotetext{a}{Archival exposure.}
\end{deluxetable}

\begin{deluxetable}{lcccccc}
\tablecaption{Properties of the NLR Clouds in NGC 1068}
\tablewidth{0pc}
\tabletypesize{\small}
\tablehead{
\colhead{Cloud} & \colhead{Area\tablenotemark{a}} & \colhead{L$_{\rm [O~III]\lambda 5007}$\tablenotemark{b}} & \colhead{Ionized H Mass}\tablenotemark{c} & \colhead{[\ion{O}{3}]$\lambda$5007 \tablenotemark{d}} &\colhead{$\log$ KE \tablenotemark{e}} & \colhead{$\log$ momentum \tablenotemark{e}} \\
\colhead{Name} & \colhead{(arcsec$^2$)} & \colhead{($10^{40}$ ergs~s$^{-1}$)} & \colhead{($10^2$ M$_\odot/n_{e,4}$)} & \colhead{/H$\beta$ flux} & \colhead{(ergs/n$_{e,4}$)} & \colhead{(dynes s/n$_{e,4}$)}}
\startdata
A   & 0.03 & \phn3.1 & \phn16.6 &  12.3 &  51.50 &  43.87 \\
B   & 0.04 & 10.8 & \phn63.0 &  11.5 &  52.33 &  44.61 \\
C   & 0.06 & 10.9 & \phn55.3 &  13.2 &  51.98 &  44.38 \\
D   & 0.04 & 21.1 & 100.8 &  13.9 &  52.18 &  44.64 \\
E   & 0.09 & 20.7 & \phn96.9 &  14.2 &  52.16 &  44.58 \\
F   & 0.06 & 24.5 & 116.2 &  13.9 &  52.58 &  44.88 \\
G   & 0.06 & \phn8.1 & \phn32.5 &  16.7 &  52.02 &  44.37 \\
H   & 0.06 & \phn3.7 & \phn16.6 &  14.7 &  51.75 &  44.11 \\
J   & 0.06 & \phn3.3 & \phn17.6 &  12.7 &  51.53 &  43.97 \\
K   & 0.06 & \phn3.5 & \phn18.8 &  12.4 &  51.65 &  44.07 \\
HV0 & \nodata & \phn3.6 & \phn20.6 &  10.9 &  51.79 &  44.09 \\
HV1 & \nodata & \phn5.0 & \phn26.4 &  11.8 &  52.17 &  44.35 \\
HV2 & \nodata & \phn1.4 & \phn\phn8.1 &  11.3 &  51.31 &  43.65 \\
HV3 & \nodata & \phn0.6 & \phn\phn5.5 &  \phn7.0 &  51.75 &  43.84 \\
HV4 & \nodata & \phn0.6 & \phn\phn4.2 &  \phn9.8 &  51.48 &  43.61 \\
HV5 & \nodata & \phn0.8 & \phn\phn5.0 &  11.2 &  51.33 &  43.60 \\
HV6 & \nodata & \phn1.1 & \phn\phn6.2 &  12.2 &  51.49 &  43.74 \\
HV7 & \nodata & \phn2.1 & \phn13.9 &  10.3 &  51.60 &  43.96 \\
\enddata
\tablenotetext{a}{Area of each distinct kinematic component on the sky, 
typically uncertain by $\pm$0.015 arcsec$^2$.}
\tablenotetext{b}{Line fluxes at velocities below galaxy systemic were 
dereddened by E(B-V)=0.35, above systemic by 0.22.}
\tablenotetext{c}{Assumes case-B recombination conditions and a pure H plasma.}
\tablenotetext{d}{Velocity averaged flux ratio, typically uncertain by $\pm0.3$.}
\tablenotetext{e}{Minimum values,
assuming that all velocity is along the line of sight.
Density n$_{e,4}$ is in units of 10$^4$ cm$^{-3}$.}
\end{deluxetable}

\end{document}